\documentstyle[prb,aps]{revtex}
\input epsf.sty
\begin{document}
%\draft
\twocolumn[\hsize\textwidth\columnwidth\hsize\csname@twocolumnfalse\endcsname

%\documentstyle[twocolumn,aps,prb,epsf,psfig,pstricks]{revtex}
%\input{epsf}
%\input epsf.sty
%\begin{document}
%\twocolumn[\hsize\textwidth\columnwidth\hsize\csname@twocolumnfalse\endcsname

\title{Signatures of Spin and Charge Energy Scales \\
in the Local Moment and Specific Heat \\
of the Two--Dimensional Hubbard Model}

\author{Thereza Paiva and R.T.~Scalettar}
\address{Physics Department, University of California, Davis, CA 95616}

\author{Carey Huscroft}
\address{Physics Department, University of Cincinnati,
Cincinnati, OH 45221--0011}

\author{A.~K.~McMahan}
\address{Lawrence Livermore National Laboratory, University of California,
Livermore, California 94550}

\date{\today}
\maketitle 

\begin{abstract}
Local moment formation driven by the on--site repulsion
$U$ is one of the most fundamental features in the Hubbard model.
At the simplest level, the temperature dependence of the local moment is
expected to have a single structure at $T \sim U$,
reflecting the suppression of the double occupancy.
In this paper we show new low temperature Quantum Monte
Carlo data which emphasize that the local moment 
also has a signature at a lower energy scale 
which previously had been thought to characterize only
the temperatures below which moments on {\it different} 
sites begin to correlate locally.
We discuss implications of these results for the
structure of the specific heat, and connections
to quasiparticle resonance and pseudogap formation in
the density of states.
\end{abstract}
\vskip2pc]

\narrowtext

\section{ Introduction}

The finite temperature properties of the two dimensional
Hubbard model have been extensively studied both
analytically and numerically.\cite{GENHUB}
Quantum Monte Carlo (QMC) is especially effective at half-filling, 
where there is no sign problem.  From calculations of the
magnetic structure factor, susceptibility,
compressibility, density of states, and
the electron self--energy, a clear picture has
emerged concerning the nature of the short and long range magnetic order,
the Mott gap, and the quasiparticle dispersion.

While the specific heat $C(T)$ has been computed by a number of
groups in one-dimension, principally
by Bethe Ansatz techniques\cite{BA},
there have been few QMC studies of $C(T)$
for the two and three dimensional Hubbard models.\cite{DUFFY,RTS1,MURA}
The behavior for large $U$ is well understood and one
expects, as in the one dimensional case, a two peak structure in $C(T)$,
with a broad high temperature peak at $T \sim U$
associated with ``charge fluctuations'', and a narrower
peak at lower temperatures associated with ``spin fluctuations''.

This two peak structure of the Hubbard model can be understood 
from a strong coupling viewpoint as follows:
At temperatures which exceed the on--site
repulsion, $T>U$, the up and down electrons are decoupled,
$\langle n_\uparrow n_\downarrow \rangle =
\langle n_\uparrow \rangle \langle n_\downarrow \rangle =\frac14$,
at half--filling, and the local moment 
$\langle m_z^2 \rangle
= \langle (n_\uparrow - n_\downarrow)^2 \rangle
= \langle n_\uparrow + n_\downarrow - 2
 n_\uparrow n_\downarrow \rangle 
 = 1 - \langle n_\uparrow n_\downarrow \rangle =0.5$,
its uncorrelated value.
At the temperature scale $T \sim U$, double occupancy begins to be 
suppressed,
$ \langle n_\uparrow n_\downarrow \rangle \rightarrow 0$,
and $m_z^2 \rightarrow 1$.
Since the potential energy in the Hubbard model
is just $P=U n_\uparrow n_\downarrow 
= {U \over 2}(1-m_z^2)$, this growth in the local
moment is synonymous with a decrease in the potential energy
as $T$ decreases,
and hence a peak in $C(T)$ at $T\sim U$.
Once moments are formed, the
half--filled Hubbard model maps onto the spin--1/2 
antiferromagnetic Heisenberg model, with exchange constant $ J = 4t^2/U$,
whose specific heat has a peak at $T \sim J$
associated with magnetic ordering.
In sum, one expects $C(T)$ for
the strong coupling Hubbard model at half--filling to have 
a ``charge peak'' at $T\sim U$ 
and a ``spin peak'' at $T\sim J$.
This strong coupling argument further suggests that the 
temperature derivatives of the potential and kinetic energies
are associated with the high and low temperature
specific heat peaks respectively.

On the other hand, the behavior of the
specific heat for small $U$ in the two dimensional Hubbard model
is still unclear.  In particular,
it is not known whether a two peak structure is present for all values of
the interaction or if these coalesce into a single peak at small $U$.
That the two peaks might merge is suggested by the fact that 
the charge and spin energy scales, $U$ and $t^2/U$, approach each other
as $U$ is decreased.  However, the situation is not so straightforward,
because the strong coupling form for the energy scale for spin ordering
crosses over to a weak coupling expression\cite{HIRSCH1,RPA}
$t \, {\rm exp}(-2 \pi \sqrt{t/U})$,
which at small $U$ is still well-separated from the energy 
scales $t$ and $U$. 

The QMC results reported in this work are consistent
with such weak coupling behavior at small $U$.  That is, one of
our key findings is that $C(T)$ shows two distinct peaks which 
persist to couplings an order of magnitude less than the
noninteracting bandwidth.
It may be that the half-filled two-dimensional Hubbard model is unique in
this respect, since, as we review below, studies in one dimension and within
DMFT show a merging of the peaks at values of $U$ roughly equal to 
the bandwidth.   Indeed, the two--dimensional half--filled Hubbard model
on a square lattice has 
unusual nesting features in its noninteracting
band structure, as well as a logarithmic 
divergence of its density of states at the Fermi level which
have previously been noted to enhance antiferromagnetism
anomalously.\cite{HIRSCH1} 

Whether this is the case or
not, there is as yet no compelling evidence of 
the appearence of a weak coupling energy scale
$t \, {\rm exp}(-2 \pi t/U)$  in the specific heat
in other dimensions. 
In one dimension, exact diagonalization in small
chains\cite{Shiba} 
and QMC calculations\cite{Schulte} suggest the two peaks merge,
but disagree with respect to the interaction strength at which this occurs.
Exact diagonalization is limited to chains
of very modest extent, and finite size effects tend
to be large for small U.  Meanwhile, QMC work
did not reach low enough temperatures  to resolve the two peaks even for large 
values of $U$ where they almost certainly both exist.
Bethe--Ansatz calculations help clarify this issue, but focus on
the large $U$ limit\cite{Usuki,Koma,Sanchez}.
Despite these various caveats, the consensus of these
approaches is that the spin and charge
peaks merge at $U/t \approx 4$, the one--dimensional bandwidth.
Quantum transfer matrix calculations \cite{QTM}
also show the merging of the two peaks at $U/t \approx 4$.

Coalescence of the specific heat peaks has also
been seen 
in ``Dynamical Mean Field Theory'' (DMFT) which studies the system in
the limit of high dimension.\cite{GEORGES,VOLLHARDT,CHANDRA} 
There, the Hubbard model
is studied with a Gaussian density of states of unit variance, and the spin
and charge peaks are found to merge at $U \approx 1.5$.
The fact that the band--width is undefined complicates comparisons 
with results in finite dimension, but one can still examine
the ratio of $U$ to the kinetic energy per particle, which
is finite for a Gaussian density of states.  
The value $U/t \approx 4$ in two dimensions has the
same ratio of $U$ to kinetic energy as the interaction strength at which
the two peak structure is lost in DMFT.
However, an additional difficulty in the interpretation of
the DMFT results, besides the use of a Gaussian density
of states, is the restriction of the calculations to the
paramagnetic phase, and therefore the neglect of antiferromagnetic
fluctuations.

% In two dimensions  Duffy and Moreo \cite{DUFFY} found that the low
% temperature peak moves to higher temperatures as U
% is decreased in weak coupling, so that the two peaks seem to be
% merging into one, as $U$ decreases.
 
It is the purpose of this paper to present a detailed study of the
temperature dependence of the local moment and the associated
features in the specific heat for the half--filled two--dimensional
Hubbard Hamiltonian.  A focus of our work will be on
extending the strong coupling picture of the two peak structure of $C(T)$ to
 intermediate and weak coupling.  As we shall show, 
the connection of moment formation
and moment ordering
with the high and low temperature peaks (respectively)
in the specific heat is modified.
At the same time, we will describe two recently 
developed techniques for computing the specific heat
which hold certain advantages over approaches
previously used.  These new techniques also allow us to
compute the entropy and free energy, 
quantities typically not so easy to obtain
with Monte Carlo.  
A fascinating conclusion of the DMFT studies\cite{GEORGES,VOLLHARDT,CHANDRA}
concerned the existence of a universal crossing point of the specific
heat curves for different $U$.  We shall show such a crossing occurs
also in two dimensions.
Finally, we will discuss results for dynamical
quantities like the density of states and optical conductivity and comment
on their consistency with the local moment and specific heat.

\section{Model and Methods}

%\vskip0.1in
%\noindent
\subsection{ The Hubbard Hamiltonian}

The two dimensional Hubbard Hamiltonian is,
\begin{eqnarray}
H &=& -t \sum_{\langle {\bf i},{\bf j} \rangle \sigma}
(c_{{\bf i}\sigma}^{\dagger} c_{{\bf j}\sigma}^{\phantom{\dagger}}
+ c_{{\bf j}\sigma}^{\dagger} c_{{\bf i}\sigma}^{\phantom{\dagger}})
\nonumber
\\
&+& U \sum_{{\bf i}} (n_{{\bf i}\uparrow} - \frac12)
(n_{{\bf i}\downarrow} - \frac12)
- \mu \sum_{{\bf i}} (n_{{\bf i}\uparrow} + n_{{\bf i}\downarrow} ).
\label{hubham}
\end{eqnarray}
Here $c_{{\bf i}\sigma}^{\dagger} (c_{{\bf j}\sigma}^{\phantom{\dagger}})$
are creation(destruction) operators for
a fermion of spin $\sigma$ on lattice site ${\bf i}$.
The kinetic energy term includes a sum
over near neighbors $\langle {\bf i},{\bf j} \rangle$ on a
two--dimensional square lattice, and the
interaction term is written in particle--hole
symmetric form so that $\mu=0$ corresponds to
half--filling $\langle n_{ {\bf i}\uparrow}+n_{ {\bf i}\downarrow}
\rangle=1$
for all Hamiltonian parameters $t,U$ and temperatures $T$.
We will henceforth set the hopping parameter $t=1$.

Equal time quantities of interest in this paper include the energy,
$E=\langle H \rangle$, the specific heat $C=dE/dT$, the
local moment $\langle m_z^2 \rangle
= \langle (n_{ {\bf i}\uparrow} - n_{ {\bf i}\downarrow})^2
\rangle$,\cite{FOOTNOTE3} and the
near neighbor spin--spin correlation function $\langle S_{{\bf i}}
S_{{\bf i}+\hat x} \rangle$.
To probe longer range magnetic order, we
evaluate the structure factor,
\begin{equation}
S({\bf Q})= {1 \over L^2} \sum_{{\bf i},{\bf j}} 
e^{i {\bf Q} \cdot({\bf i} -{\bf j})} \langle
(n_{{\bf i}\uparrow}-n_{{\bf i}\downarrow})
(n_{{\bf j}\uparrow}-n_{{\bf j}\downarrow})\rangle ,
\end{equation}
where ${\bf Q}=(\pi,\pi)$ is the antiferromagnetic wave vector. 

We also evaluate two dynamic quantities.
The density of states $N(\omega)$
is given implicitly from QMC data
for the imaginary time Green's function, 
\begin{equation}
G(\tau)= {1 \over N} \sum_{{\bf p}}
\langle c({\bf p},\tau) c^{\dagger}({\bf p},0) \rangle
=\int_{-\infty}^{+\infty} d \omega
{e^{- \omega \tau} N(\omega)
\over
e^{-\beta \omega}+1  } \, .
\label{dos}
\end{equation}
Likewise, the optical conductivity, $\sigma_{xx}(\omega)$,
is related to QMC data
for the imaginary time current--current correlation function,
\begin{eqnarray}
\sigma_{xx}(\tau)&&= 
\langle j_x(\tau) j_x(0) \rangle
=\int_{-\infty}^{+\infty} d \omega
{e^{- \omega \tau} 
\sigma_{xx}(\omega)
\over
e^{-\beta \omega}-1  } \, ,
\nonumber
\\
j_x(\tau) &&= -it \sum_{{\bf i}\sigma}
(c_{{\bf i}+ \hat x,\sigma}^{\dagger} c_{{\bf i}\sigma}^{\phantom{\dagger}}
-c_{{\bf i}\sigma}^{\dagger} c_{{\bf i}+ \hat x,\sigma}^{\phantom{\dagger}}).
\label{opticon}
\end{eqnarray}
Both $N(\omega)$ and $\sigma_{xx}(\omega)$ are computed using the
Maximum Entropy (ME) technique\cite{MEM}
to invert the integral relations.

%\vskip0.1in
%\noindent
\subsection{ Determinant Quantum Monte Carlo}

We use determinant QMC\cite{DQMC} to evaluate the expectation values
above.  This approach treats the electron--electron correlations
exactly, and at half--filling, where we focus this work, is able to
produce results with very small statistical fluctuations
at temperatures low enough that the ground state  has been reached.
The technique is limited to finite size lattices, and we
will show appropriate scaling analyses to argue that
we extract the thermodynamic limit.

\subsection{ Calculation of Specific Heat}

We will evaluate the specific heat in three ways\cite{FOOTNOTE2}.
All begin by using QMC to obtain $E_n=E(T_n)$ and
the associated error bars $\delta E_n$
at a sufficiently fine grid of $N_T$
discrete temperatures $T_n$.
The first  approach is
straightforward numerical differentiation
of the energy $E(T_n)$.  The second utilizes a fit to the
numerical data for the energy $E(T_n)$,
and the  third is an approach using the ME method
to invert the data $E(T_n)$ to obtain a spectrum of  excitations
of the system.
These last two approaches were introduced relatively
recently.\cite{HUSCROFT1,HUSCROFT2}
Therefore we shall describe them in some detail.

In our fitting method, whose results we denote
by $E_e(T)$,  we match the QMC data $E_n$
to the functional form,\cite{HUSCROFT1}
\begin{equation}
E_e(T)=E(0)+\sum_{l=1}^{M} c_{l} e^{-\beta l \Delta} \, ,
\label{Ee}
\end{equation}
by adjusting the parameters  $\Delta$ and $c_l$ 
to minimize,\cite{bevington}
\begin{equation}
\chi^2 = {1 \over N_T} \sum_{n=1}^{N_T} { (E_e(T_n) - E_n))^2 \over 
(\delta E_n)^2 } \,.
\label{chi2}
\end{equation}
The number of parameters $M$ is chosen to be about one--fourth of the
number of data points to be fit.
Smaller numbers do not allow a good fit, while larger ones overfit the
data.  We  find that a range of intermediate $M$ exists 
which gives stable and consistent results.

Calculation of $C(T)$ by fitting $E(T)$ to
polynomials has also been used recently,\cite{DUFFY}
but requires at least two separate functions to be used
at high and low temperatures.
An advantage of Eq.~5 is that it uses a
single functional form over the entire $T$ range, and
has the correct low and high temperature limits,
$C(T) \rightarrow 0$.

The specific heat can also be evaluated 
by differentiating an expression which relates the 
energy to the density of states
of Fermi and Bose excitations in the system,\cite{HUSCROFT2}
\begin{eqnarray}
E_{{\rm me}}(T)&=&-\int_{-\infty}^{+\infty}
d\omega \hskip0.1cm
\omega [ F(\beta,\omega) \rho_F(\omega)
+B(\beta,\omega) \rho_B(\omega) ]
\nonumber
\\
 F(\beta,\omega) &=&  {1 \over 1 + e^{\beta \omega} }
\hskip0.8cm
 B(\beta,\omega) =  {1 \over 1 - e^{\beta \omega} },
\label{Eme}
\end{eqnarray}
and differentiating to get the specific heat,
\begin{eqnarray}
C_{{\rm me}}(T)&&={\partial E_{{\rm me}}(T) \over \partial T}
\nonumber
\\
=-\int_{-\infty}^{+\infty}
d\omega \hskip0.1cm &&
\omega [ {\partial F(\beta,\omega) \over \partial T}
\rho_F(\omega)
+
{\partial B(\beta,\omega) \over \partial T}
\rho_B(\omega)].
\label{Cme}
\end{eqnarray}
The integral equation for $E_{{\rm me}}(T)$ is inverted
by using the ME method
to obtain $\rho_F(\omega)$ and $\rho_B(\omega)$ 
from the QMC data for $E_n$.
We denote by $E_{\rm me}(T)$ the energy
obtained from the  resulting 
$\rho_F(\omega)$ and $\rho_B(\omega)$.

This ME approach differs in philosophy from 
the  fitting approach which begins with a physically reasonable 
functional form $E_e(T)$ and then minimizes the deviation $\chi^2$
from the numerical data.
Instead, ME computes the most probable spectrum $\rho(\omega)$ given
the energy data $E(T_n)$ and kernals $F(\beta,\omega)$
and $B(\beta,\omega)$,
{\it without} presupposing a particular functional form.
Despite this difference, we will show that the results of 
the two techniques are very similar, and agree quite well
with numerical differentiation.

\subsection{ The Entropy and Free Energy}

Both the ME and fitting techniques
allow the specific heat, entropy, and free energy 
to be computed by the standard formulae, 
\begin{eqnarray}
C(T)&=&{dE(T) \over dT},
\nonumber
\\
S(T)&=&\int_{0}^{T} {C(T') \over T'} dT',
\nonumber
\\
F(T)&=&E(T)-TS(T).
\label{C}
\end{eqnarray}
Here $E(T)=E_e(T)$ or $E_{\rm me}(T)$.

In the case of the fitting technique,
we can evaluate the sum rule,
\begin{equation}
{1 \over N}
\int_{0}^{\infty} dT {C(T) \over T} =\sum_{l=1}^{M} {c_l \over l \Delta}
=  2 \, {\rm ln}\,  2 -S_0 \, ,
\label{Cfit}
\end{equation}
which ties the high temperature entropy to the logarithm of the
dimension of the Hilbert space.
The $T=0$ entropy $S_0$ must of course vanish in the
thermodynamic limit.\cite{S0}
For the 2--d Hubbard model at $U \neq 0$ we find
the term $S_0$ vanishes even on finite lattices, but it may be present in
other Hamiltonians.
In the present work, the sum rule of Eq.~10 is satisfied to a few percent.
For the maximum entropy method,
a similar check is possible by integrating $\rho_{F}$.

We now turn to the results of our simulations.

\section{Equal Time Correlations-- Local Moment, Specific Heat,
and Magnetic Order}

\subsection{The Local Moment}

In the introduction we reviewed the standard argument for
the expected behavior of the Hubbard model local moment.
Early determinant QMC work for the 2--dimensional Hubbard model\cite{HIRSCH1}
confirmed this,
as did subsequent investigations.\cite{WHITE2}
In Fig.~1 we show that an examination of $m_z^2$ with a
fine temperature mesh and at low temperatures reveals that 
after reaching a plateau at intermediate temperatures,
$m_z^2$ changes value again at a second, low temperature, scale.\cite{EARLIER}
We will come back to this point in more detail later, but it is worth
commenting immediately that while the low temperature structure in
$m_z^2$ is small compared to the size of the growth at high temperature,
it occurs over a much smaller temperature range, and hence contributes a
large peak in the specific heat.

In order to determine whether this is a finite size effect,
in Fig.~2 we show data on a range of lattice sizes from 4x4 to 10x10. 
The evidence for the existence of the low energy scale 
is robust as the lattice size is increased.
We can also make an extrapolation to the thermodynamic limit
assuming a correction which goes as the inverse of the linear
system size, as spin--wave theory 
indicates is appropriate for the full structure factor.\cite{SWT}

\begin{figure}[hbt]
\epsfxsize=8cm
\begin{center}
\epsffile{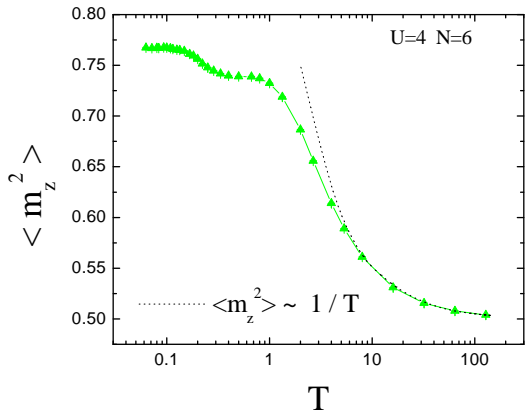}
\caption{
The temperature dependence of the local moment is 
shown at fixed $U=4$ for a 6x6 lattice.  In addition to rising
at $T \sim U$, as $T$ decreases,
$m_z^2$ exhibits a second structure at lower temperature.
}
\end{center}
\end{figure}

\begin{figure}[hbt]
\epsfxsize=8cm
\begin{center}
\epsffile{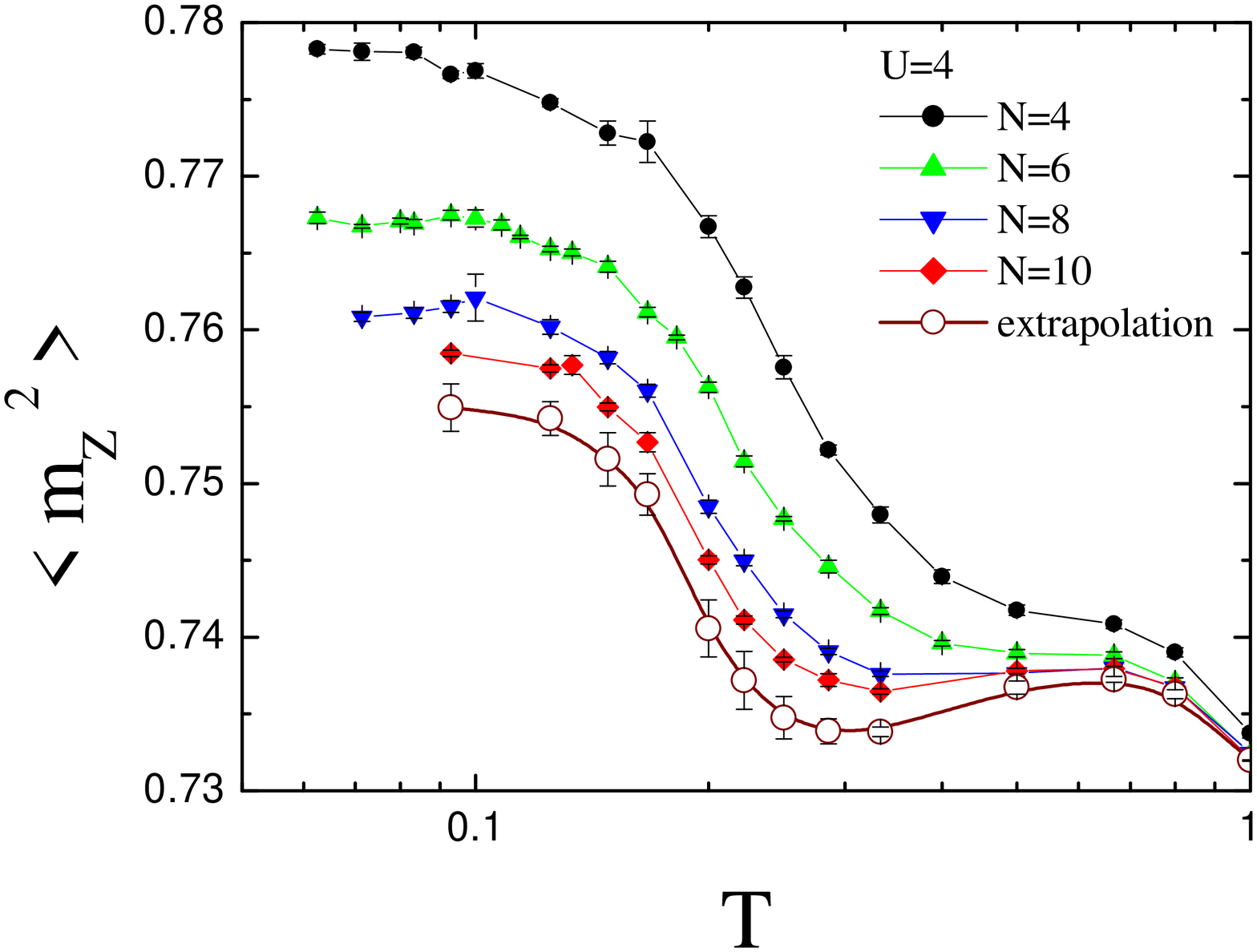}
\caption{
Data for the local moment for different lattice sizes.
The low temperature feature in $m_z^2$ remains present
as the lattice size is increased.
}
\end{center}
\end{figure}

\begin{figure}[hbt]
\epsfxsize=8cm
\begin{center}
\epsffile{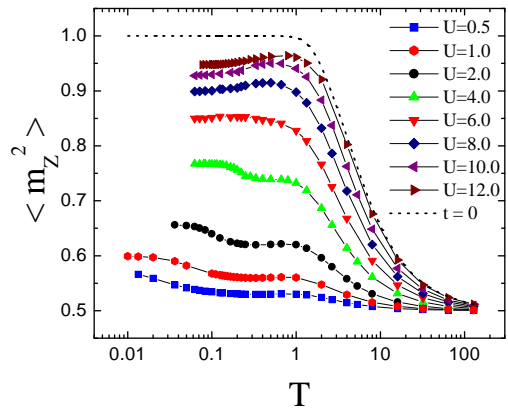}
\caption{
Temperature dependence of the local moment for different values of the
on--site repulsion $U$ on 6x6 lattices. The dashed line corresponds to the
zero-hopping limit, for $U=12$.
The position of the low temperature feature first increases
in temperature as $U$ increases, but then gradually falls, as emphasized 
later in Fig.~\ref{peakpos}.
}
\label{mz2}
\end{center}
\end{figure}

\begin{figure}[hbt]
\epsfxsize=8cm
\begin{center}
\epsffile{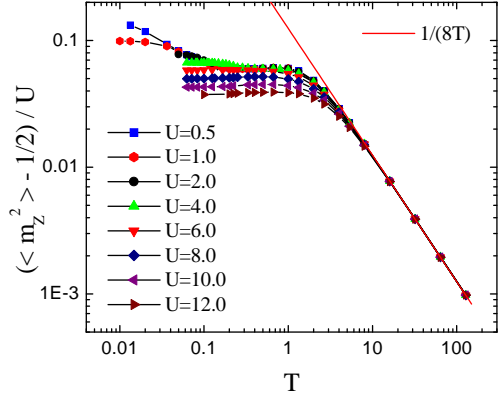}
\caption{
Scaled version of the data of Fig.~\ref{mz2}.
The plot shows that for high $T$ the local moment
$m_z^2 = \frac12 + {U \over 8T}$, that is, the deviation of the
local moment from its noninteracting value $\frac12$ 
exhibits a universal behavior with a temperature scale $U$.
}
\label{mz2scaled}
\end{center}
\end{figure}

Additional insight is obtained by looking at the behavior of
$m_z^2$ at different values of $U$, as in Fig.~\ref{mz2}.
The data of Fig.~\ref{mz2} are replotted in Fig.~\ref{mz2scaled} to emphasize
the universal
nature of the high temperature behavior and the fact that the 
initial increase in the local moment as temperature is decreased does
indeed occur at 
a temperature scale set by $U$.

\begin{figure}[hbt]
\epsfxsize=8cm
\begin{center}
\epsffile{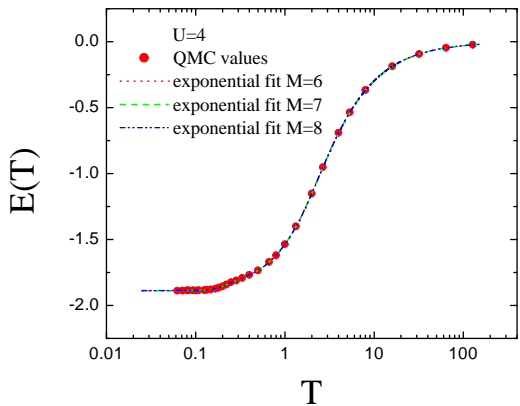}
\caption{
QMC values (circles) for the energy, and  fittings  
provided by Eq.~5.  Here $U=4$ and the lattice is 6x6.
The different lines (dotted, dashed, dot-dashed) show that the fit is stable
over a range of
values of the number of fitting parameters.
}
\label{energy}
\end{center}
\end{figure}

In the zero-hopping ($t=0$) limit, $\langle m_z^2 \rangle= 
1/ \{ {\rm exp}[-U/(2T)]+1 \}$, 
which is monotonic with $T$, dropping from 1 at $T=0$ to 1/2 
at $T \rightarrow \infty$. Why is the
maximum in $\langle m_z^2 \rangle$ shifted from $T=0$ 
when the hopping is present?
The ground state is antiferromagnetic, with the effective 
exchange $J$ arising from virtual hopping of the electrons. 
This virtual transfer reduces 
the degree of localization as can be seen from the 
values of $\langle m_z^2 \rangle$ at $T=0$ in 
Fig.~\ref{mz2}. In the low-lying excited states the deviations from the 
antiferromagnetic state reduce the virtual hoppings since
the Pauli principle forbids hopping when adjacent electron spins are 
ferromagnetically aligned.
Localization is thereby increased with 
increasing temperature, giving rise to the maximum at 
$T \ne 0$. This maximum at large $U$ 
has also  been observed in  one-dimension.\cite{Shiba}

%What is the physical interpretation of the low energy scale
%in the temperature dependence of the moment seen in Fig.~3, at strong
%coupling, large U? %At half--filling, 
%the local moment is related to the double occupancy 
%$D=\langle n_\uparrow n_\downarrow \rangle$ by
%$\langle m^2 \rangle 
%= 1 - 2 D$.
%Thus a maximum in 
%$\langle m^2 \rangle $
%corresponds to a minimum in $D$ which can be interpreted as
%resulting from
%the free energy gain with localizing the particles
%(decreasing $D$) as $T$ is increased.\cite{GEORGES}
%An alternative way to view the reduction in the moment which occurs
%at low temperatures is by connecting it to the behavior of the density 
%of states.  It is now well established that as $T$ is lowered
%a quasiparticle resonance develops in the density of 
%states.\cite{GEORGESL,DMFT2,QMCHM}
%This peak shifts spectral weight out of the lower and upper
%Hubbard bands, which are associated with well defined moments,
%into a feature with a (small) free--electron--like moment.
%As a consequence, the total moment is reduced.

At weak coupling, we see the opposite effect.  The local 
moment has instead an additional increase at low temperature.
This has a natural explanation in terms of the formation of
local magnetic order.  If there is an energy gain with
ordering, there will be an associated preference for large moments.
It is interesting to note that the 
DMFT results\cite{GEORGES} do not observe this additional moment
enhancement at weak coupling.  Instead, the local moment always has
a maximum as a function of temperature.  This is, perhaps,
a consequence of restricting the DMFT to the paramagnetic phase.

\subsection{ The Energy and Specific Heat}

\begin{figure}[hbt]
\epsfxsize=8cm
\begin{center}
\epsffile{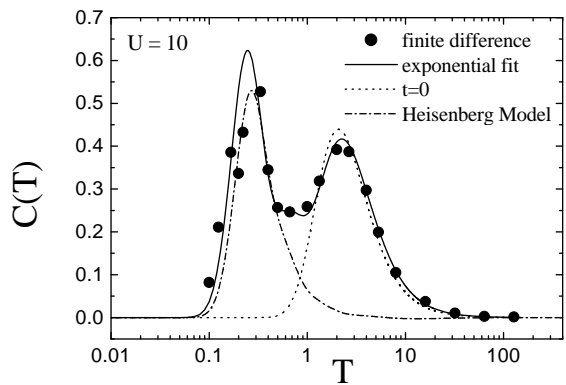}
\caption{
Results for the specific heat at $U=10$ and lattice size 6x6.
The  circles are numerical differentiation and the solid line is
from the exponential fitting approach.
The dotted line is the result for $C(T)$ at $t=0$ (that is, a single site).
The dashed line is $C(T)$ for the Heisenberg model with
$J=4t^2/U=0.4$.
}
\label{ctu10}
\end{center}
\end{figure}

Figure \ref{energy} shows
the QMC results for $E(T_n)$ together with the
exponential fit $E_e(T)$.
% Polynomials, which have previously been used\cite{DUFFY}
% require two separate fits for low and ``high'' temperatures, and,
% in fact, would break down at the highest temperatures where
% they cannot reporduce $C(T\rightarrow \infty)=0$
% which is required for systems with an energy
% spectrum bounded from above.
Calculation of the specific heat brings out the
low temperature features in $E(T)$.
We begin our analysis of this data for  the specific heat by looking
at the data at relatively strong coupling ($U=10$) as shown in
Fig.\ \ref{ctu10}.  It is seen that the results for the Hubbard model
are beautifully fit by combining the zero-hopping $t=0$ specific heat,
which lies
right on the high $T$ Hubbard model results,
and the Heisenberg specific heat,\cite{LOOP} which similarly lies 
right on the low $T$ Hubbard model results.
The areas under both the low and high $T$ peaks are precisely
$\ln{2}$, as expected for the high $T$ loss of entropy associated with
moment
formation and then low $T$ moment alignment.
Clearly, this provides a good understanding of the strong
coupling specific heat, as well as demonstrates the reliability of
our approach to computing $C(T)$.

\begin{figure}[hbt]
\epsfxsize=8cm
\begin{center}
\epsffile{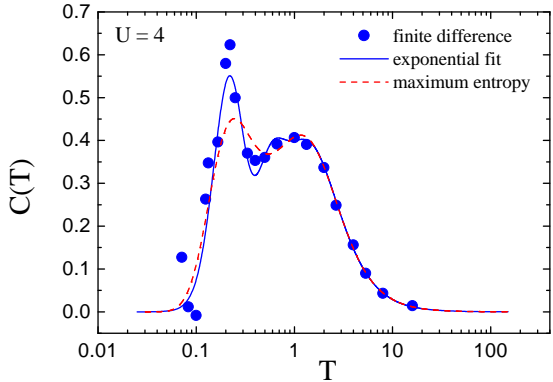}
\caption{
Results for the specific heat at $U=4$ and lattice size 6x6.
The exponential fitting (full line) and ME results 
(dashed line) are in good agreement
with direct numerical differentiation of the QMC data.
Both smooth out the noise associated with direct numerical 
differentiation of the QMC data, though ME appears to broaden the results
perhaps a bit too much.
}
\label{ctu4}
\end{center}
\end{figure}

\begin{figure}[hbt]
\epsfxsize=8cm
\begin{center}
\epsffile{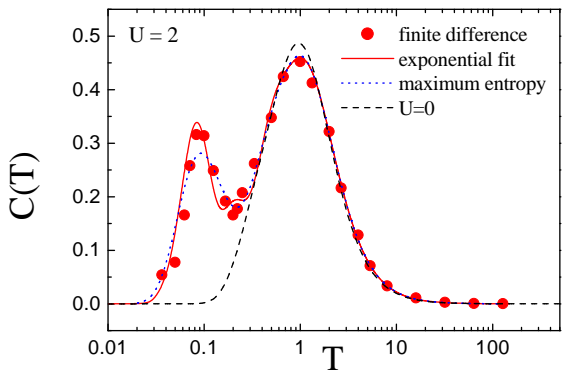}
\caption{
Results for the specific heat at $U=2$ and lattice size 6x6.
The exponential fitting (full line) and ME results 
(dotted line) are both in good agreement
with direct numerical differentiation of the QMC data.
As at $U=4$, ME produces somewhat broader peaks.
The dashed line is the specific heat for the
non-interacting limit ($U=0$).  
}
\label{ctu2}
\end{center}
\end{figure}

Further confirmation of the accuracy of our $C(T)$ calculations is
evident by comparing results for the numerical differentiation, the 
 exponential fit,  and the ME techniques as shown  in
Figs.\ \ref{ctu4} and \ref{ctu2}.  It is seen that the agreement between
all
three approaches is good.  Figs.\ \ref{ctu4} and \ref{ctu2} also emphasize
that
both the exponential fit and ME techniques are well suited to capturing the
two energy scales in the problem.
It is also interesting to comment on the $U$ dependence
of the areas under the specific heat curves of Fig.~\ref{ctu10},
\ref{ctu4}, 
and \ref{ctu2}.
 As $U$ decreases into the weak coupling regime,
the low $T$ peak has less and less entropy.
For $U=2$ the area is only about $\ln{2}/2$.
This will be discussed at greater length shortly.

\begin{figure}[hbt]
\epsfxsize=8cm
\begin{center}
\epsffile{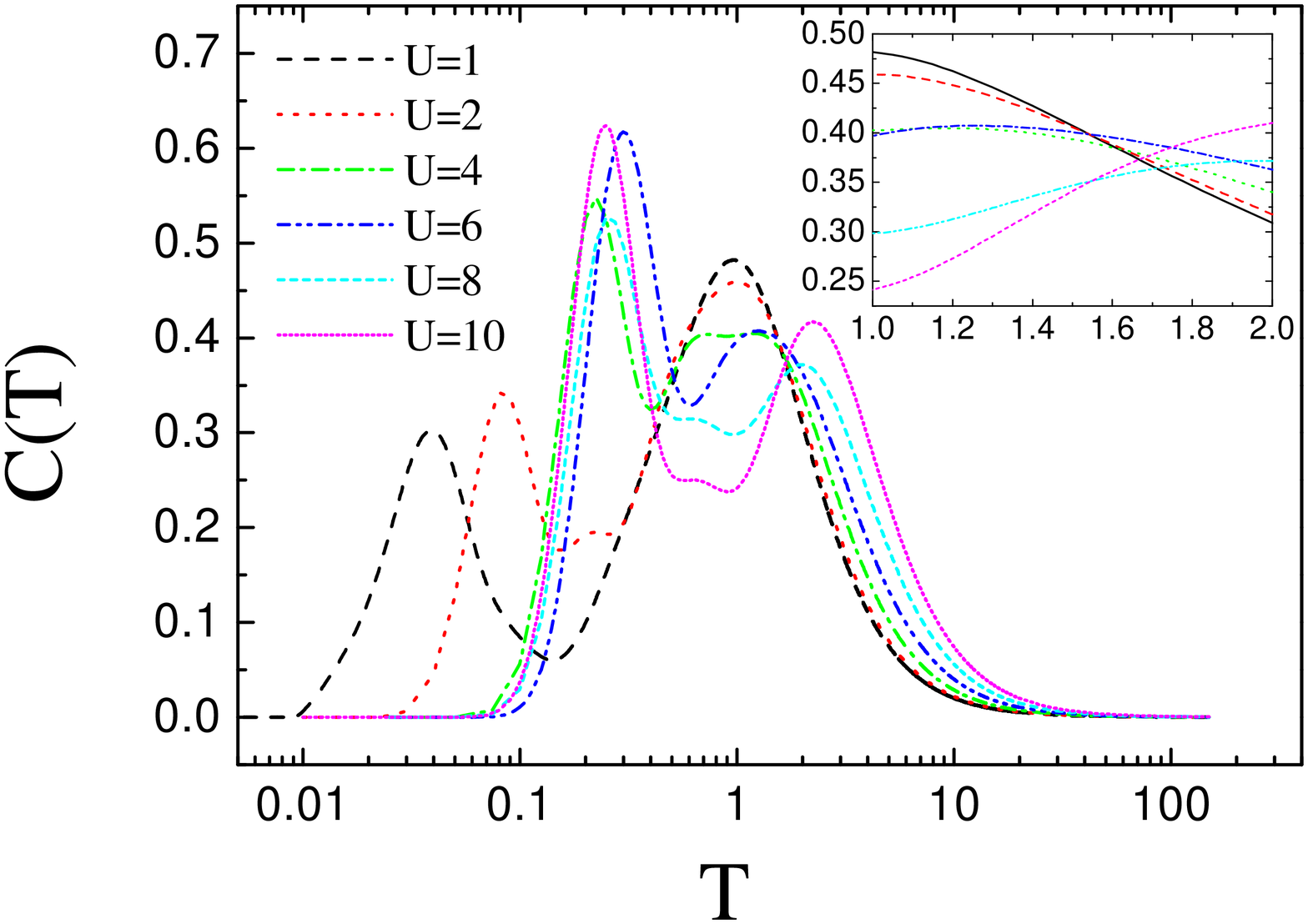}
\caption{
The specific heat  curves for different $U$ values show a nearly universal
high temperature crossing for the two
dimensional Hubbard model, as has previously been observed in DMFT
and in several experimental systems.
}
\label{crossing}
\end{center}
\end{figure}

When the specific heat curves for different $U$ are plotted together,
as in Fig.~\ref{crossing}, one sees that there is a nearly universal
crossing at high temperature.
There has been considerable recent discussion of this 
phenomenon, both its occurrence in experimental systems like 
$^3$He and heavy fermion systems,
and models like the Hubbard Hamiltonian.\cite{GEORGES,VOLLHARDT,CHANDRA}
In the case of the Hubbard model, 
the crossing has been argued to follow from the fact that
the high temperature entropy is independent of $U$,
ln4 = $\int_{0}^{\infty} C(T,U) dT/T,$
which implies that  $0 = \int_{0}^{\infty}  \partial C/ \partial U dT/T$.
Hence $ \partial C/ \partial U$ must be positive for some temperature ranges and
negative for others, a condition for crossing to occur.\cite{VOLLHARDT}
The narrowness of the crossing region is traced ultimately to
the linear temperature dependence of the double occupancy,
the conjugate variable associated with $U$.\cite{VOLLHARDT}
In DMFT, two crossings were observed for the Hubbard model, with the
high temperature one being nearly universal, while the low
temperature intersections were considerably more spread out.

Previous studies in two dimensions\cite{DUFFY}
exhibit crossings of the specific heat at
$T_*=1.6$, with a crossing region $\Delta T_*=0.2$.  
%More
%but with a considerably wider crossing region
%$\Delta T_*=0.2$.  
%Moreover,
%but in that case  the crossings shift systematically
%to larger temperature as $U$ is increased.
In Fig.~\ref{crossing} we confirm this result that
a specific heat crossing occurs in two dimensions.
While the crossings in the earlier study\cite{DUFFY}
shift systematically with $U$, we instead see 
a random fluctuation of the crossing point.
This suggests that the width of the crossing 
we report here is
%might be considerably lessened by data with tighter error bars, that it may be 
dominated by statistical fluctuations as opposed
to possible systematic effects.
We have also verified that the double occupancy
has a linear temperature dependence at low $T$, especially at
weak coupling, which is the criterion established for a universal
crossing point.

\begin{figure}[hbt]
\epsfxsize=8cm
\begin{center}
\epsffile{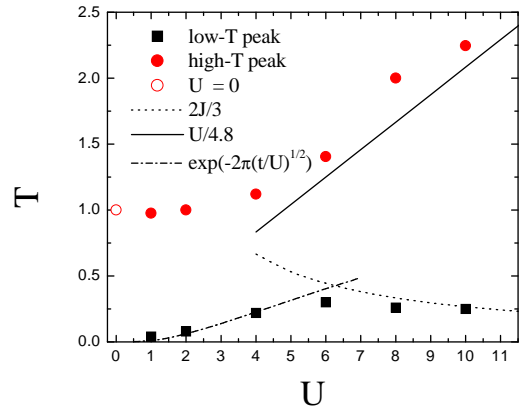}
\caption{
Position of the high $T$ and low $T$ peaks of the specific heat. The
dotted line is the Heisenberg limt $T \sim 2J/3$, the full line
corresponds to the $t=0$ limit, $T \sim U/4.8$ and the dash-dotted 
line corresponds to an RPA--like form for the temperature scale of
the antiferromagnetic spin fluctuations,
$T \sim {\rm exp}[-2 \pi \sqrt{t/U}]$. 
}
\label{peakpos}
\end{center}
\end{figure}

The position of the two peaks as a function of $U$ is shown in 
Fig.~\ref{peakpos}.
The strong coupling analysis gives us firm predictions for the
peak positions at large $U$:  First, the $t=0$ result tells us 
that the high $T$ peak is at $T_{{\rm high}} \approx U/4.8$.  The 
deviation from this limit seen in Fig.~\ref{peakpos} can be ascribed to
quantum fluctuations. 
Meanwhile, the Heisenberg result for the low $T$ peak is at
$T_{{\rm low}} = 2J/3=8t^2/3U$. \cite{jaklic}
We similarly understand the position of
$T_{{\rm high}}$ at weak coupling from the $U=0$
analysis:  $T_{{\rm high}} \approx t=1$ in units where $t=1$.
The value of $T_{{\rm low}}$ for weak coupling is somewhat more problematic.
In three dimensions, the Ne\'el temperature which describes the
onset of long--range magnetic order,
has a non--monotonic behavior with $U$,\cite{RTS1,DMFT}
first rising\cite{RPA} at small $U$ as
$T_N \propto {\rm exp}[-2 \pi t/U]$
and subsequently falling back down as
$T_N \propto t^2/U$ at large $U$.
In lower dimensions, like the 2--d case studied in this
paper, $T_N=0$. Nevertheless, in weak coupling,
both the random phase approximation\cite{RPA} and Hartree-Fock
calculations
give a finite $T_N \propto {\rm exp} [-2\pi \sqrt{t/U}]$ in 2--d. It is
tempting in this case to interpret this energy scale as that of the short
range spin fluctuations which give rise to the low temperature peak in $C$
at weak coupling, and similarly $t^2/U$ as the corresponding energy scale
at strong coupling. Indeed, both the increase for small $U$, consistent
with the exponential form, and the subsequent decrease can be seen in
$T_{{\rm low}}$ in Fig. ~\ref{peakpos}. It might also be noted that the
entropy in the low-$T$ Hartree-Fock $C(T)$ peak goes to $0$ as
$U\rightarrow 0$, which is also consistent with the decreasing entropy
under the low-$T$ QMC $C(T)$ peak as $U$ becomes
small.\cite{LOWTLOWUENTROPY}

\begin{figure}[hbt]
\epsfxsize=8cm
\begin{center}
\epsffile{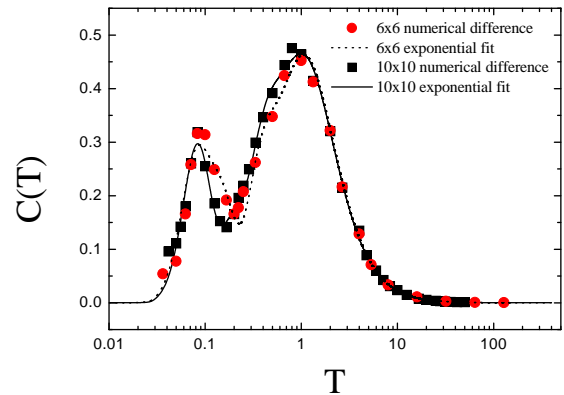}
\caption{
Comparison of 
results for the specific heat at $U=2$ and lattice sizes 6x6 and 10x10.
}
\label{fse}
\end{center}
\end{figure}

Finite size effects are illustrated in Fig.~\ref{fse}, which shows the
data for the specific heat, obtained by finite differention
of the energy data, on 6x6 and 10x10 lattices at $U=2$.
As is seen, the error bars as inferred from the
scatter in the data are of the same size as any possible
systematic effect.
In determinant QMC, finite size effects are largest at
weak coupling, so this data represents a rather stringent test
of possible lattice size dependence of our results for the
thermodynamics.

We now turn to the issue of the separate contributions of
the kinetic $K$ and potential $P$ energies to the specific heat.
As discussed in the introduction, we might associate the charge peak
with the potential energy, since the energy $U$ is what enforces
double occupancy and reduces charge fluctuations.
Since the energy scale $J = 4t^2/U$ arises from 
virtual hopping processes,
it is more naturally associated with the kinetic energy.
At strong coupling this 
division describes the mapping of the Hubbard model
onto the Heisenberg model, and then the specific  heat of the
Heisenberg model itself, and works extremely well quantitatively,
as seen in Fig.~\ref{ctu10}.

At intermediate coupling it is not
so natural to consider separately the derivatives
of the kinetic and potential energies, as these
quantities mix dramatically.
Indeed, the behavior of the local moment shown in Fig.~1
indicates that at $U=4$ the potential energy in fact
contributes both to the low and high temperature
structure of $C(T)$.

\begin{figure}[hbt]
\epsfxsize=7cm
\begin{center}
\epsffile{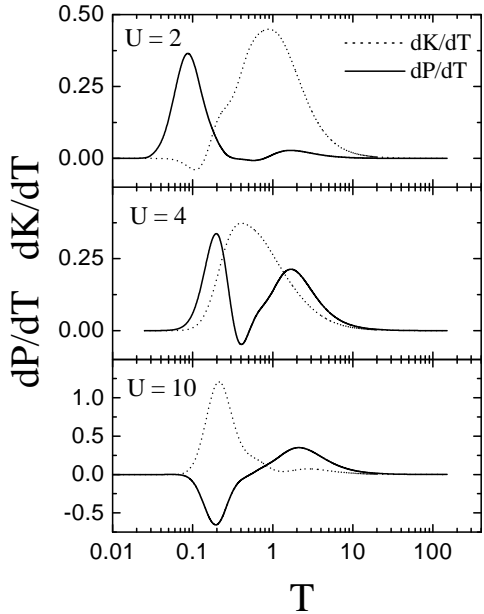}
\caption{
The separate temperature derivatives of the kinetic and potential energy.
At strong coupling, the temperature dependence of the potential energy
and kinetic energies give rise to the 'charge' and spin peaks in $C$ 
respectively.  This attribution is interchanged at weak coupling.
}
\label{dkdt}
\end{center}
\end{figure}

Figure \ref{dkdt} shows $dP/dT$ and $dK/dT$ for $U=2,4,10$.
At strong coupling $U=10$, $dP/dT$ has a high temperature
maximum, while $dK/dT$ has a low temperature peak, as expected.
In the combined specific heat, then, $P$ and $K$ are responsible
for the charge and spin peaks respectively.
Interestingly, however, even at large $U$, $dP/dT$ 
has a significant negative dip
at low $T$, reflecting the potential energy cost of delocalization.
As we have remarked,
this effect has previously been noted in the 
1--d Hubbard model and DMFT.\cite{Shiba,GEORGES}

The positions of the contributions of $dP/dT$ and $dK/dT$
to $C(T)$ are exchanged as $U$ is decreased.  Finally,
at $U=2$, it is the potential energy which is responsible
for the low temperature 'spin' peak, and the kinetic energy for
the high temperature 'charge' peak.

\begin{figure}[hbt]
\epsfxsize=7,5cm
\begin{center}
\epsffile{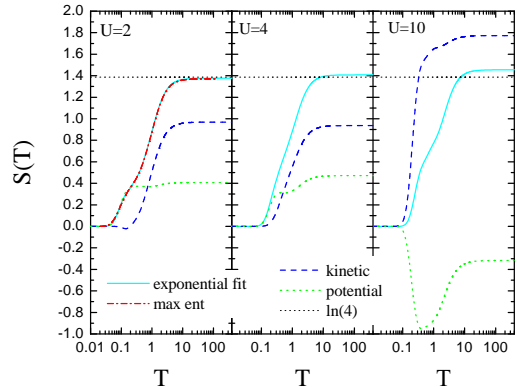}
\caption{
The temperature dependence of the entropy.
}
\label{entropy}
\end{center}
\end{figure}

The entropy $S$  can also  be obtained as the area
under $(1/T) C(T)$, as well as the separate kinetic
and potential contributions.  Fig.~\ref{entropy}
shows the results at $U=2$,  $U=4$ and $U=10$. In all cases the value of
$S$ at high temperature equals the expected $2\ln{2}$ to  within
a few percent.  
At strong coupling, $U=10$, the offsetting kinetic and potential
contributions seen in the low-$T$ peak region in Fig.~\ref{dkdt} are
reflected as well in Fig.~\ref{entropy}. Nevertheless, the total entropy
(solid curve) shows a shoulder at $\ln{2}$ and then the final
high-temperature value of $2\ln{2}$ reflecting the two peaks in
Fig. \ref{ctu10}, which correspond first to the ``Heisenberg'' disordering
of the spins and then at higher temperature to the destruction of local
moments.
At weak coupling, the initial increase in
entropy at low $T$ comes from the temperature dependence of
the potential energy which, as we have seen,
is what gives rise to the low $T$ peak in $C(T)$.  
The area under the low $T$ peak in $C(T)/T$
is reduced from its large $U$ value of $\ln{2}$.
Certainly one origin of this decrease is the reduction of
the local moment 
$m_z^2$ from its large $U$ value $m_z^2=1$ 
at small $U$, as seen in Fig.~\ref{mz2}.
The entropy associated with local ordering of 
moments scales with the moment size.\cite{LOWTLOWUENTROPY}

\subsection{ Magnetic Correlations}

\begin{figure}[hbt]
\epsfxsize=8cm
\begin{center}
\epsffile{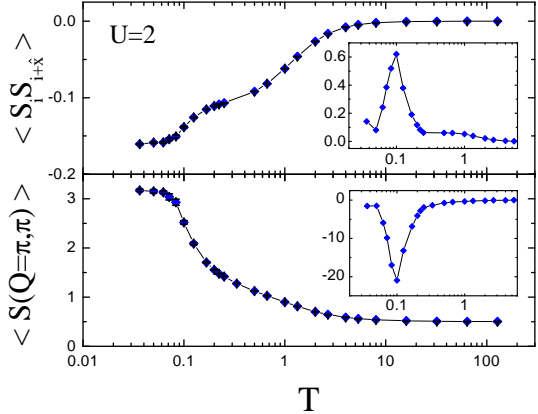}
\caption{
Spin-spin correlation between neighboring sites and magnetic structure factor
as a function of temperature for a 6x6 lattice with $U=2$. The insets show
the derivatives of these two quantities with temperature. 
}
\label{sisi1}
\end{center}
\end{figure}

It is natural to associate the low temperature feature in
the local moment and therefore the low $T$ peak in $C(T)$ with  the onset
of {\it local} antiferromagnetic 
correlations between neighboring spins. 
To understand how these correlations develop  we have calculated the
spin-spin correlation function between
neighboring sites  $\langle S_i S_{{\bf i}+\hat x} \rangle$ 
and  the magnetic structure factor $S({\bf Q})$.
Figure \ref{sisi1} shows  these two quantities as a function of
temperature for $U=2$. The inset shows the derivatives of the spin-spin
correlation function for neighboring sites 
($d S_{\bf i} S_{{\bf i}+\hat x}/dT$) and for
the magnetic structure factor ($d S({\bf Q}) /dT$) with respect to the
temperature: the 
sharp peaks  form roughly at the same position  as the
specific heat has its low $T$ peak. Since $T_N=0$, the peak in $d S({\bf
Q}) /dT$ should not be associated with long ranged correlations.
Indeed, the largest contribution to $d S({\bf
Q}) /dT$ comes from antiferromagnetic correlations between neighboring
spins.

\section{Density of States and Optical Conductivity}

\subsection{ The Density of States:}

While the standard interpretation of the two peak structure
in the specific heat in terms of the freezing out of charge and
spin fluctuations, at high and low $T$ respectively,
is consistent with our data at large $U$,
Fig.~\ref{dkdt} emphasized that such a picture is not as useful
at weak coupling.

\begin{figure}[hbt]
\epsfxsize=8cm
\begin{center}
\epsffile{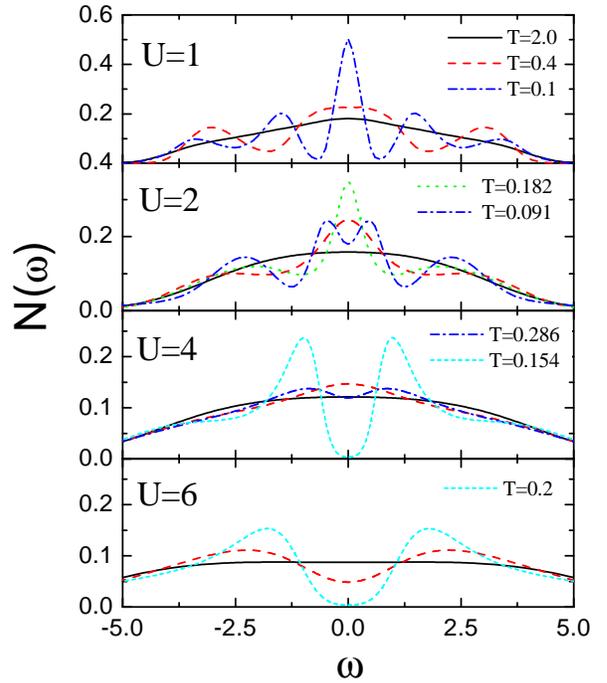}
\caption{
The temperature evolution of the
density of states at $U=1, 2, 4, 6$ on a 6x6 lattice.
For $U=1,2$, as $T$ is lowered, the single broad peak first evolves into a
sharper quasiparticle peak before a pseudogap opens.
The quasiparticle peak is washed out as $U$ increases.
}
\label{nomega}
\end{center}
\end{figure}

Greater insight into the physics behind the specific heat is obtained by
looking at the dynamics.  Figure \ref{nomega} shows 
results for the density of states $N(\omega)$
for $U=1, 2, 4, 6$ and decreasing temperatures.
At high $T$ the density of states consists of a  
single, very broad, bump  with maximum   at $\omega=0$.
As $T$ is lowered for the smaller $U$ values (e.g., $U=2$),
$N(\omega=0)$ first increases as 
a quasiparticle peak develops at  $\omega=0$.
This peak appears to be very similar to that found
in multiband models like the periodic Anderson model, 
where  it is associated with a Kondo resonance.
As $T$ is lowered yet further, $N(\omega)$ then begins to decrease as a
dip begins to form in the center of the quasiparticle peak. 

\begin{figure}[hbt]
\epsfxsize=8cm
\begin{center}
\epsffile{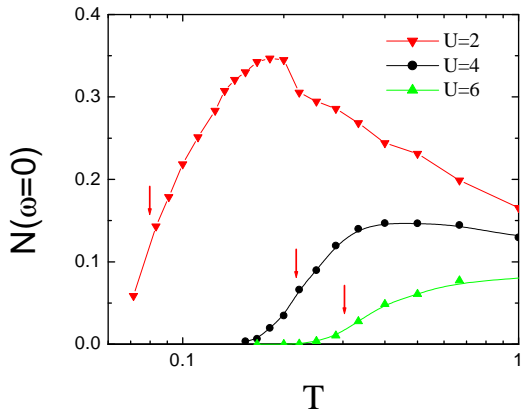}
\caption{
The zero frequency density of states at $U=2, 4, 6$ and 6x6 lattice.
The arrows indicate the location of the low $T$ peaks in the specific heat.
}
\label{nw0}
\end{center}
\end{figure}

For larger
$U$, on the other hand (e.g., $U=6$), only the dip develops with
decreasing
temperature, and so $N(\omega=0)$ always decreases as $T$ is lowered. 
This behavior is emphasized in
Fig.~\ref{nw0} which shows $N(\omega=0)$ as a function of temperature
for $U=2, 4, 6$.
The temperature at which $N(\omega=0)$ becomes small is seen in
Fig.~\ref{nw0} to increase with $U$, and it appears
correlated with the position of the low $T$ peak in $C(T)$.
 Indeed, the derivative of $N(\omega=0)$ is maximum at 
the same temperature where $C(T)$ has its 
low $T$ peak  (indicated by arrows in the plot).

This ``pseudogap'' in the density of states is one of the central features 
of the 2--d Hubbard model under recent discussion,
since it is one of the most interesting features of the
normal state of the high temperature superconductors
in the underdoped regime.\cite{PSEUDOGAPEXPT}
As the pseudogap has a d--wave symmetry, like the superconducting
order parameter itself, it is believed
to arise as a result of short--range spin fluctuations
which might also play a role in the pairing.
The pseudogap's existence has a long history of discussion,
and debate, in the numerical literature on the Hubbard model
which we shall now review, since concerns about possible finite size 
effects in the pseudogap may bear on similar concerns in
the behavior of the specific heat.

The dynamical data presented in  Figs.~\ref{nomega}--\ref{nw0} 
do not make a fully compelling case for the 
relation between the low $T$ peak in $C(T)$ 
and the pseudogap, since they 
are for a single fixed lattice size.
A particular issue is
the behavior of the 
density of states $N(\omega)$ of the half--filled
Hubbard model in the thermodynamic limit.  Quantum Monte Carlo
results concerning this question are still evolving.  Early simulations
had the somewhat surprising conclusion that 
a pseudogap in $N(\omega)$ was present
at weak to intermediate coupling
only at $T=0$ in the thermodynamic limit.
That is, while on a fixed lattice size $L$ a gap in $N(\omega)$
develops at a finite temperature $T$, it would go away 
if the lattice size were increased.\cite{WHITE1}
Meanwhile, at strong coupling, the same work found the pseudogap 
persists at finite $T$ even as the system size increases.
This behavior was interpreted as reflecting the
fact that long range antiferromagnetic
correlations are present only at $T=0$, and that
such long range correlations were required for a gap in $N(\omega)$
for small $U$.
This interpretation was questioned, however, since
one might expect the pseudogap to depend only on the existence of
short range antiferromagnetic correlations.
Such local order should form at a temperature which is independent of
lattice size, leading to the conclusion that
the pseudogap should be present below that temperature even
on large lattices.

If the original suggestion that the finite temperature
pseudogap disappears at weak coupling 
in the thermodynamic limit were the case, 
it might raise similar questions about possible finite size effects in
our  results for
the low  temperature  structure of the magnetic moment and specific heat.
We believe this is not a concern for three reasons.
First,
one can consider the limit of very weak coupling.  As we already see in 
Fig.~\ref{ctu2}, the high $T$ peak in $C(T)$ is well fit by a noninteracting 
calculation, and specifically therefore comes from the kinetic energy.
If the local moment (potential energy) did {\it not} evolve
at low $T$, then $C(T)$ would have a single peak structure.
Therefore,
our separate finite size scaling analysis 
for the moment and the specific heat support  each other.
Second,
we have presented data at different finite sizes (Fig.~\ref{fse}) which show
no
evidence for the low $T$ peak shifting with increased lattice size.
Finally, recent work suggests that the pseudogap 
exists in the thermodynamic limit at weak coupling and is not
a finite size effect there.\cite{TREMBLAY1,TREMBLAY2,HUSCROFTP,FOOTNOTE4} 

% In any case, it is evident that the association of the
% low $T$ peak in $C(T)$ at weak coupling with pseudogap formation
% is not inconsistent with recent data on the behavior of the
% pseudogap on finite lattices.

\subsection{ The Optical Conductivity}

The density of states itself does not present a complete picture
of the nature of the excitation gap.  A more refined view may be
obtained by looking at the optical conductivity, as shown in 
Figs.~17--19. 
These results show that the Hubbard model has a non--zero charge gap,
even at weak to intermediate coupling where $U$ is less than the bandwidth.
We examine the dynamic spin susceptibility (not shown) and find  the spin
gap
to vanish since the 2--d Hubbard model has long range magnetic order
at $T=0$ and hence ungapped, power law, spin wave excitations.

\begin{figure}[hbt]
\epsfxsize=8cm
\begin{center}
\epsffile{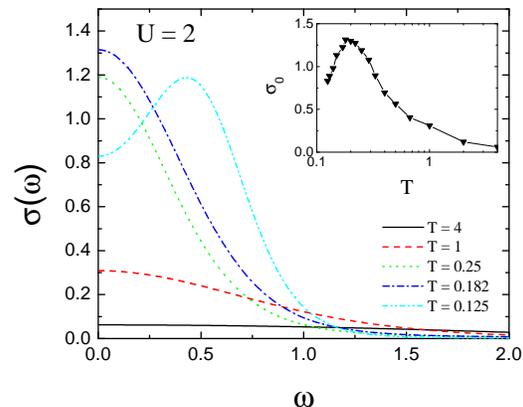}
\caption{
The frequency dependence of the optical conductivity for $U=2$ and
different temperatures.
Inset:  The zero frequency value as a function of temperature, indicating
the opening of a Mott gap.
}
\end{center}
\end{figure}

\begin{figure}[hbt]
\epsfxsize=8cm
\begin{center}
\epsffile{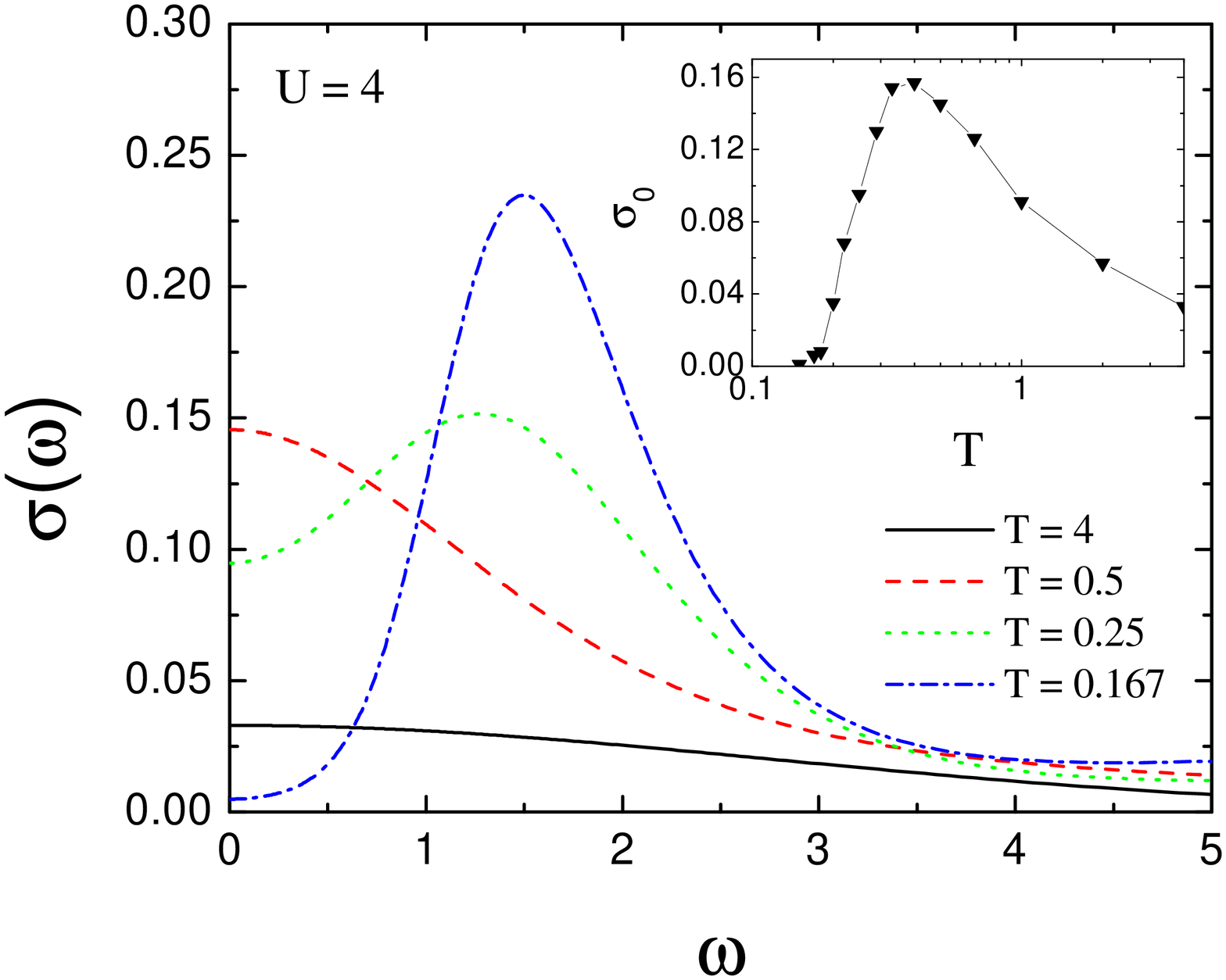}
\caption{
Same as Fig.~17 except $U=4$.
}
\end{center}
\end{figure}

\noindent
\begin{figure}[hbt]
\epsfxsize=8cm
\begin{center}
\epsffile{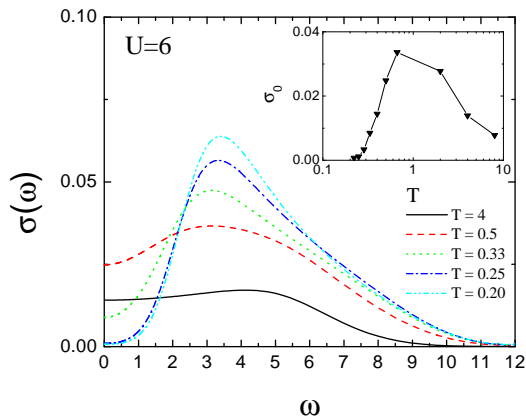}
\caption{
Same as Fig.~17 except $U=6$.
}
\end{center}
\end{figure}

\section{Conclusions}

In this paper we have examined carefully
the low temperature structure of the 
local moment and specific heat of the two dimensional Hubbard model.
A striking conclusion of our work is that the two peak structure in
the specific heat is preserved down to $U = t = W/8$,
where $W=8t$ is the bandwidth.  In one dimension, the coalescence
of the spin and charge peaks appears to occur at much larger 
$U$, namely the one dimensional bandwidth.  
Meanwhile, in infinite dimension, 
the peaks also come together at an interaction strength associated with
an average kinetic energy per particle corresponding to
half the two dimensional bandwidth.
However, we have pointed out that the argument that the peaks come
together which is based on a comparison of the 
scales $4t^2/U$ and $U$ might be flawed, 
as the weak coupling spin energy scale is instead set by
$t \, {\rm exp}(-2 \pi \sqrt{t/U})$.
This leaves open the question of why the separation of
the specific heat energy scales is maximal in intermediate dimension.
Possibly the half-filled two-dimensional Hubbard model is unique due to the
unusual characteristics of its noninteracting states at the Fermi
energy.\cite{HIRSCH1} However, it should also be recalled that the
DMFT studies have been restricted to the paramagnetic phase, which may have
an important impact on the existence of two well defined peaks. Finally
the existence of the Nagaoka state in two dimensions, but not in one or
infinite  dimensions, indicates that there may be no reason to expect
systematic behavior here as a function of dimension.\cite{SchlottmannPC}

We have emphasized that
the standard nomenclature which identifies the high temperature
peak in the specific heat as due to ``charge'' fluctuations
and the low temperature peak
as due to ``spin'' fluctuations
while useful at strong coupling,
needs to be refined.  At weak coupling, the high temperature peak comes
from the kinetic energy
while the low temperature peak comes from the potential energy.
By comparing with the behavior of the density of states,
we have argued that the structure of $C(T)$  may be associated with 
``pseudogap'' formation, that is, the
onset of short range antiferromagnetic correlations between
near--neighbor spins.

A detailed understanding of the relationship of the energy and 
local moment formation is desirable in a number of contexts in
using model Hamiltonians to describe strongly correlated materials.
In particular, while minimization of the energy determines the 
dominant low temperature phases,
the various types of behavior of the local moment 
(for example screening by conduction electrons in multi--band models)
can also provide important clues concerning the suitability of
different models in describing the low  temperature physics.\cite{HELD1}

\vskip0.2in
\noindent
\acknowledgments
Work at UCD was supported by the CNPq-Brazil, the
LLNL Materials Research Institute, and
NSF--DMR--9985978;
that at LLNL, by
the U.S. Department of Energy under Contract No.~W--7405--Eng--48.
We thank K.~Held, M.~Jarrell, M.~Martins, P.~Schlottmann, and M.~Ulmke for
useful
discussions.


\begin{references}
\bibitem{FOOTNOTE3}
We actually evaluate all three components of the local moment
$m_x^2$, $m_y^2$, $m_z^2$.  They are, of course, equal.
\bibitem{FOOTNOTE2}
In classical Monte Carlo, the best approach to evaluate $C(T)$
is often to use the fluctuation formula
$C= \beta^2 (\langle E^2 \rangle - \langle E \rangle^2) $.
This is not a useful way to proceed for determinant
Quantum Monte Carlo because the square of the
Hamiltonian involves products of six and eight fermion operators,
and the associated expectation values have large fluctuations.
\bibitem{EARLIER}
Our QMC data agree precisely with earlier work, which, however, did not go
to quite so low temperatures and also used a more coarse temperature grid,
and hence apparently did not detect the second low $T$ structure.
\bibitem{LOWTLOWUENTROPY}
Even though the low $T$ peak in $C(T)$ is not arising from long
range magnetic
order, the $U$ dependence of its area bears a resemblance to what happens at
$T_N$ in a Hartree--Fock (HF) calculation.  
In HF, and in the thermodynamic limit, both paramagnetic
and antiferromagnetic $C/T$ have the full $2\ln{2}$ area.  The difference
is that as $T$ is reduced below $T_N$, the antiferromagnetic (AF)
$S(T)$ drops rapidly to zero,
while the PM $S(T)$ drops gradually to zero.  In HF these are second order
transitions, so $S(T)$ is continuous, $S_{{\rm PM}}(T_N)=S_{{\rm AF}}(T_N)$.  
But there is a
discontinuity for the derivative, $C(T)/T$.  So going up in $T$ from $T=0$
the AF $C(T)$ stays ~0 (AF gap) while the PM $C(T)$ is larger and grows
larger.  Then just before $T_N$ the AF $C(T)$ shoots up way above the PM
$C(T)$, as $T$ first accesses the DOS to either side of the AF gap,
followed by the discontinuity in $C(T)$ dropping back to the PM value at
$T_N$. This has the appearance of a peak, which gets rounded off
for finite sizes.  Since $S_{\rm PM}(T_N)=S_{\rm AF}(T_N)$ and 
$T_N, S_{\rm PM}(T_N) \rightarrow 0$  as  $U\rightarrow 0$ it
is clear that the area under this peak also goes to 0.
\end{references}
\end{document}